\documentclass[aps, prl, twocolumn, superscriptaddress, 10pt, nofootinbib]{revtex4-1}
\usepackage[sort&compress]{natbib}   
\usepackage{amsmath,amssymb,amsfonts}       
\usepackage{graphicx}                       

\usepackage{xcolor}                          
\usepackage[colorlinks=true]{hyperref}    
\usepackage[UKenglish]{babel}
\usepackage[UKenglish]{isodate}
\usepackage{multirow}
\usepackage{braket}

\usepackage{enumitem}

\newcommand{\im}{\mathrm{i}}
\newcommand{\gtwo}{\mathit{g}^{(2)}}
\newcommand{\magic}{BNS}

\begin{document}
	\title{Raman Quantum Memory with Built-In Suppression of Four-wave Mixing Noise}
	\author{S. E. Thomas$^{1,2}$, T. M. Hird$^{1,3}$, J. H. D. Munns$^{1,2}$, B. Brecht$^{1,4}$, D. J. Saunders$^{1}$, J. Nunn$^{5}$, I. A. Walmsley$^{1,2}$, P. M. Ledingham}
	
	\affiliation{Clarendon Laboratory, University of Oxford, Parks Road, Oxford, OX1 3PU, UK\\
		$^2$QOLS, Department of Physics, Imperial College London, London SW7 2BW, UK \\
		$^3$Department of Physics and Astronomy, University College London, London WC1E 6BT, UK \\
		$^4$Applied Physics, University of Paderborn, Warburgerstrasse 100, 33098 Paderborn,Germany \\
		$^5$Centre for Photonics and Photonic Materials, University of Bath, Bath, BA2 7AY, UK}
	
	\date{\today}
	
	\begin{abstract}
		Quantum memories are essential for large-scale quantum information networks. Along with high efficiency, storage lifetime and optical bandwidth, it is critical that the memory add negligible noise to the recalled signal. A common source of noise in optical quantum memories is spontaneous four-wave mixing. We develop and implement a technically simple scheme to suppress this noise mechanism by means of quantum interference. Using this scheme with a Raman memory in warm atomic vapour we demonstrate over an order of magnitude improvement in noise performance. Furthermore we demonstrate a method to quantify the remaining noise contributions and present a route to enable further noise suppression. Our scheme opens the way to quantum demonstrations using a broadband memory, significantly advancing the search for scalable quantum photonic networks.
	\end{abstract}
	
	\maketitle

	\emph{Introduction} -- An optical quantum memory is a device that can faithfully store and release quantum states of light on demand. This is a key element for future photonic quantum information protocols as a means to synchronise probabilistic processes via multiplexing, enabling secure long-distance communication through the distribution of entangled states~\cite{Lvovsky2009}. Many impressive implementations of on-demand QM protocols have been demonstrated across various platforms including warm~\cite{EIT2005, Reim2011, Wolters2017, Namazi2017,Zugenmaier2018,Kaczmarek2018,Guo2019} and cold atomic vapours~\cite{Dudin2013, Ding2015,Cho2016, ATS2017, Hsiao2018,Vernaz-Gris2018} and solid-state systems~\cite{CRIB2010, Hedges2010, Heinze2013, England2015, Seri2017}. While considerable progress has been made to reach high efficiencies~\cite{Hedges2010, Cho2016, Hsiao2018} and long storage times~\cite{Dudin2013,Heinze2013}, designing a memory protocol that does not add noise to the recalled signal remains a significant challenge. Common noise processes include atomic resonant fluorescence, spontaneous Raman scattering (SRS) from unpumped thermal population from the storage state, and spontaneous four-wave mixing (SFWM)~\cite{Heshami2016}. Eliminating these without sacrifice to the memory lifetime or storage bandwidth has proven difficult. These spurious noise processes pollute the desired memory output field, severely limiting the signal to noise ratio (SNR), which in turn upper-bounds the achievable fidelity for storing and recalling qubits~\cite{Gundogan2015}, as well as significantly modifying the photon number statistics i.e. the second order auto-correlation function $\gtwo$~\cite{England2015,Michelberger2015}. Fluorescence can be reduced by operating off-resonance, and SRS from the storage state can be eliminated by near-perfect optical pumping. However removing SFWM intrinsic to broadband quantum memories remains the final hurdle.

	Here we propose and demonstrate a technically simple method to suppress four-wave mixing noise in atomic memories. Our method is widely applicable to any atomic species, since it harnesses the strong linear absorption of the atomic ensemble itself. Romanov et al. have shown that SFWM noise can be suppressed via Raman absorption of the noise photon into a second atomic isotope~\cite{Romanov2016,Prajapati:17}. Inspired by their protocol we present a new scheme for noise suppression, by operating the memory at a specific detuning from resonance such that the unwanted noise field is resonant with the populated atomic transition. In this arrangement, the competing processes of noise generation by SFWM, and its absorption and dispersion by the atomic resonance significantly suppress anti-Stokes scattering and thus the contamination of the signal field. By operating the atomic memory at a specific detuning we utilise this built-in noise suppression (\magic) mechanism, achieving significant reduction of noise without any detrimental effects on the memory efficiency or lifetime, nor any change to the memory initialisation. This is in contrast to other, more complex, strategies to suppress SFWM noise such as cavity engineering~\cite{Saunders2016}, polarisation selection rules~\cite{Zhang2014}, non-colinear geometry~\cite{Dabrowski2014a}, and phase-mismatching in dispersive media~\cite{England2015}. We also present a new method to analyse different noise processes in quantum memories and quantify how these affect the photon number statistics of the retrieved state, thus enabling us to give a recipe for the remaining steps necessary for high fidelity retrieval in broadband quantum memories. Therefore, our scheme holds great promise as a route towards a technically simple, noise-free quantum memory.
	
	\begin{figure}
		\includegraphics[width=0.9\linewidth]{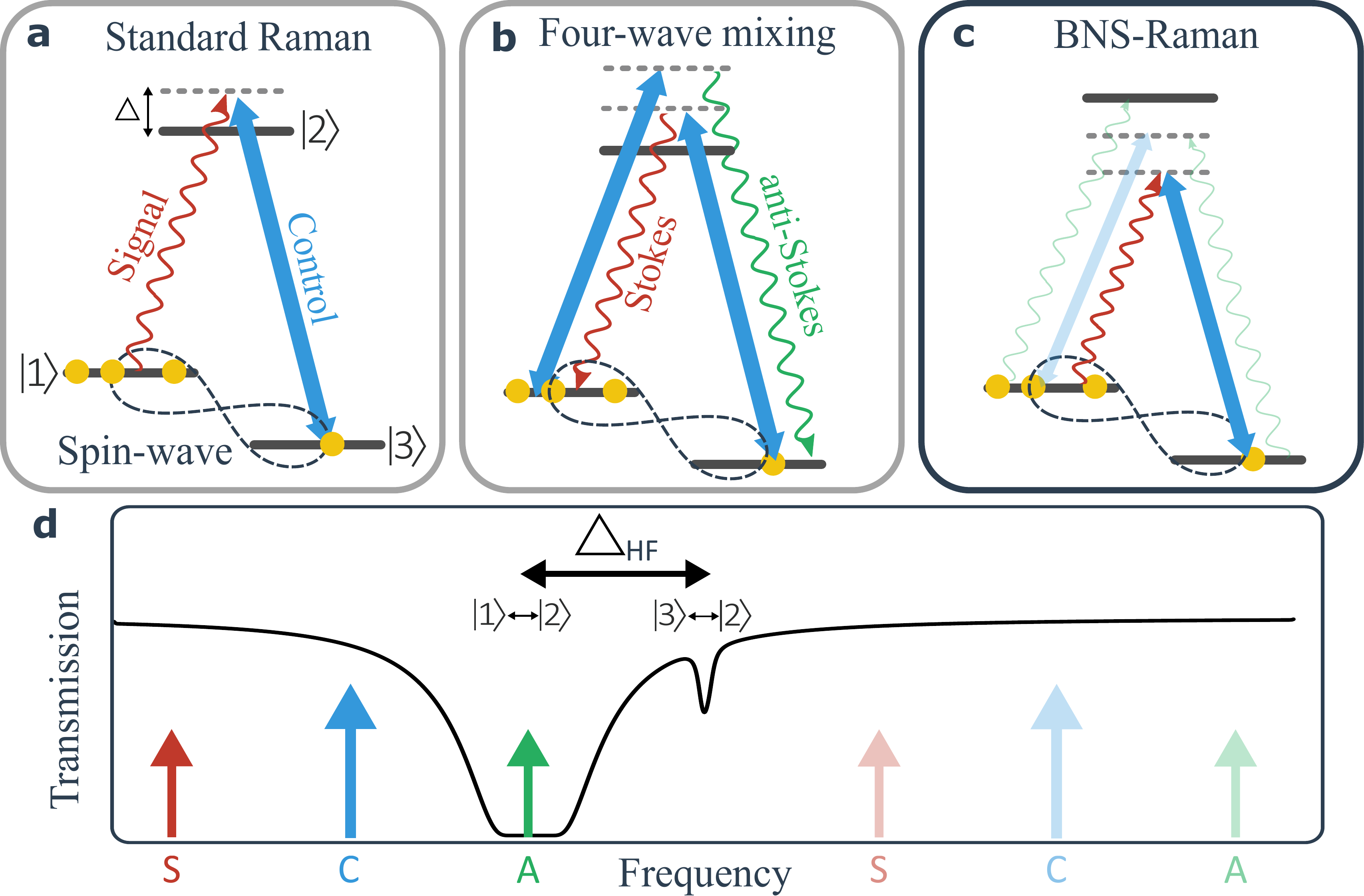}
		\caption{ 	{(a) The Raman Memory protocol uses a strong control field to drive a Raman transition and store an input signal field as a coherence across an atomic ensemble, or spin-wave. (b) SFWM noise arises when the control field couples to the populated state $\ket{1}$ and drives anti-Stokes scattering. (c) The absorptive Raman memory operates at $\Delta = - 2\Delta_\mathrm{hf}$ such that the anti-Stokes field is on resonance with the $\ket{1} \rightarrow \ket{2}$ transition, and anti-Stokes scattering is strongly suppressed. (d) Simulated absorption spectrum of warm caesium vapour in a nitrogen buffer gas at $83^\circ$C, with 99.9\% of the population in state $\ket{1}$. The arrows show the frequency of the signal/Stokes, control and anti-Stokes fields for the \magic\,(left) and STD (right) Raman protocols.} \label{fig:Schematic}}
	\end{figure}
	
	\emph{\magic\,Raman Protocol} -- The built-in noise suppression scheme is investigated in an off-resonant Raman memory in warm caesium vapour. The Raman memory protocol is based on an ensemble of atoms each with a $\Lambda$-energy level configuration, which are initialised in the long-lived ground state, $\ket{1}$ (see Fig.~\ref{fig:Schematic}a). We apply a strong control pulse in two-photon resonance with a weak signal to drive a stimulated two-photon Raman transition from $\ket{1}$ to $\ket{3}$, while the two fields themselves are detuned from the excited state $\ket{2}$ by $\Delta$. This coherently stores the signal field as a collective excitation, or spin-wave, across the entire ensemble of atoms. To retrieve the signal, a second control pulse is applied which drives the reverse process and coherently converts the atomic spin-wave back into an optical field. However, this protocol inherently suffers from SFWM noise which inhibits operation at the quantum level~\cite{Michelberger2015}. The origin of SFWM noise is the unwanted coupling of the strong control field to the populated ground state, $\ket{1}$, which drives spontaneous anti-Stokes scattering (see Fig.~\ref{fig:Schematic}b). This creates a \emph{noisy} spin-wave that has significant overlap with the \emph{memory} spin-wave, and which is efficiently read out with the same temporal and spectral mode as the signal field. This adds thermal noise to the retrieved photonic state, preventing quantum storage.
	
	By operating the Raman memory with the control field at a detuning of $\Delta_s = -2 \Delta_\mathrm{hf}$, we ensure that the anti-Stokes field is on resonance with the $\ket{1} \rightarrow \ket{2} $ transition and undergoes strong linear absorption. This is a coherent process and interferes with the SFWM generation process, since both fields are at the same frequency. The destructive interference of these two pathways suppresses the generation of noise photons at the signal frequency. One may interpret this mechanism physically as the continual absorption of anti-Stokes photons that are generated by SFWM. The absorption length for the anti-Stokes field in this configuration is inversely proportional to the optical depth of the ensemble, $d$, and is typically sub~100~$\mu$m. Any anti-Stokes photon that is scattered will be absorbed within a characteristic length scale $\sim L/{d}$, where $L$ is the length of the atomic ensemble, and the generated spin-wave excitation is therefore localised. Thus SFWM cannot lead to a delocalised excitation over the entire atomic ensemble, in contrast to the read-in and read-out memory interactions. The spatial confinement of the noisy spin-wave means that the overlap with the memory spin-wave mode is greatly diminished. This process is akin to the dissipative quantum Zeno effect~\cite{Popkov2018}, in that the absorption of a noise photon acts as a measurement process which prevents noise excitations collectively building up over the memory interaction length.
	
	\emph{Results} -- A schematic of the experimental setup is shown in Figure~\ref{fig:ExptSetup} and details are given in the Supplemental Material. We operate the memory in two configurations: (1) $\Delta_s = -2 \Delta_\mathrm{hf}$ where the anti-Stokes field is on resonance with the atomic transition and therefore strongly absorbed (\magic-Raman); (2) $\Delta_s = +2 \Delta_\mathrm{hf}$, where the coupling strength of the Raman memory interaction is the same but there is no atomic suppression of the SFWM noise i.e. the ``standard'' Raman memory (STD-Raman). 
	
	\begin{figure}
		\centering
		\includegraphics[width=\linewidth]{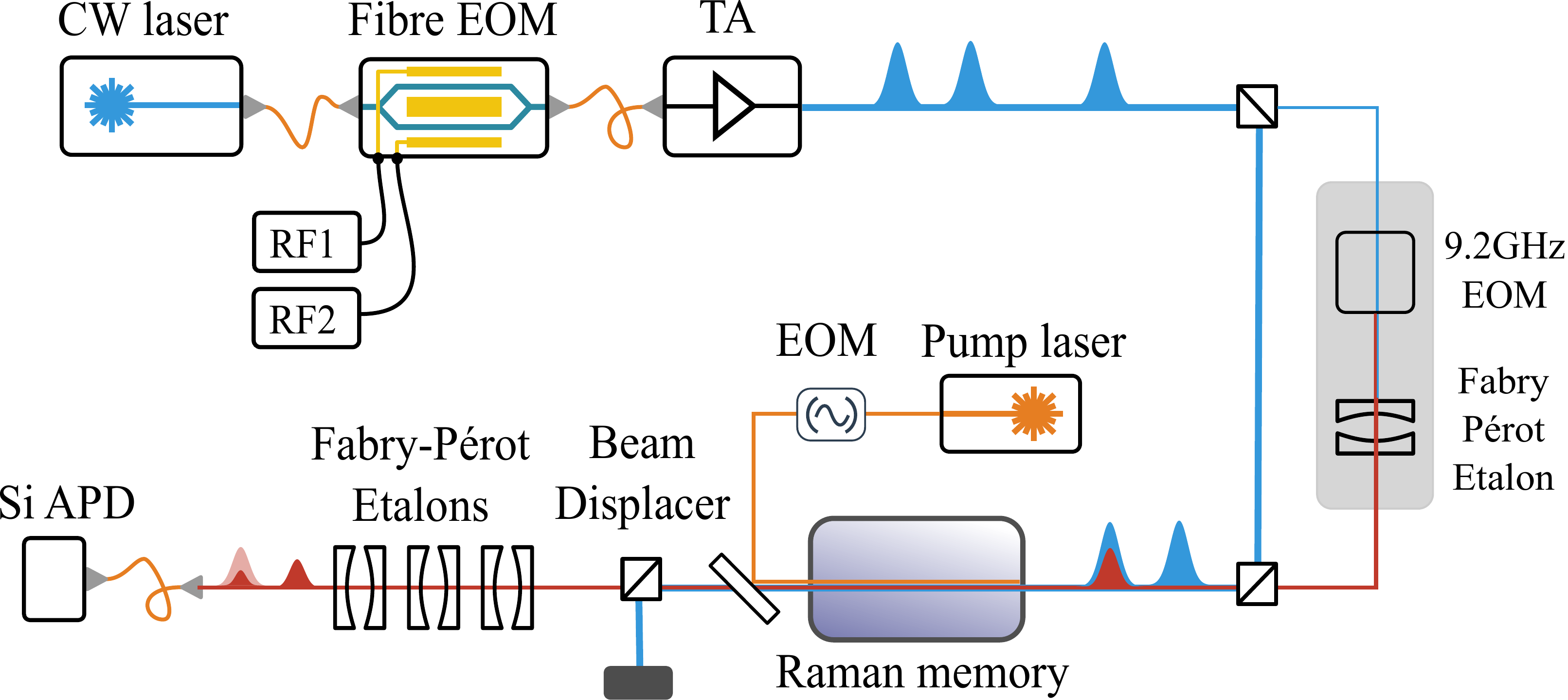}
		\caption{{Experimental setup of the Raman memory. The signal and control pulses are carved from a CW laser using a fibre integrated electro-optic interferometer (Fibre EOM) driven by RF signals from two arbitrary waveform generators. The pulses are amplified using a tapered amplifier (TA) and split into two arms. The signal pulses pass through an EOM driven at 9.2~GHz to generate a sideband at the signal frequency, and delayed in time to overlap with the strong control pulses. The pulses interact with an ensemble of warm caesium atoms, which is heated to 83$^\circ$C and initialised in $\ket{1}$ by a counter-propagating optical pumping laser. After the memory the control pulse is filtered from the signal using a beam displacer and Fabry-P\'{e}rot etalons, before the signal is detected using a silicon single-photon avalanche photodiode (Si APD).} \label{fig:ExptSetup}}
	\end{figure}

\begin{figure}
		\includegraphics[width=0.85\linewidth]{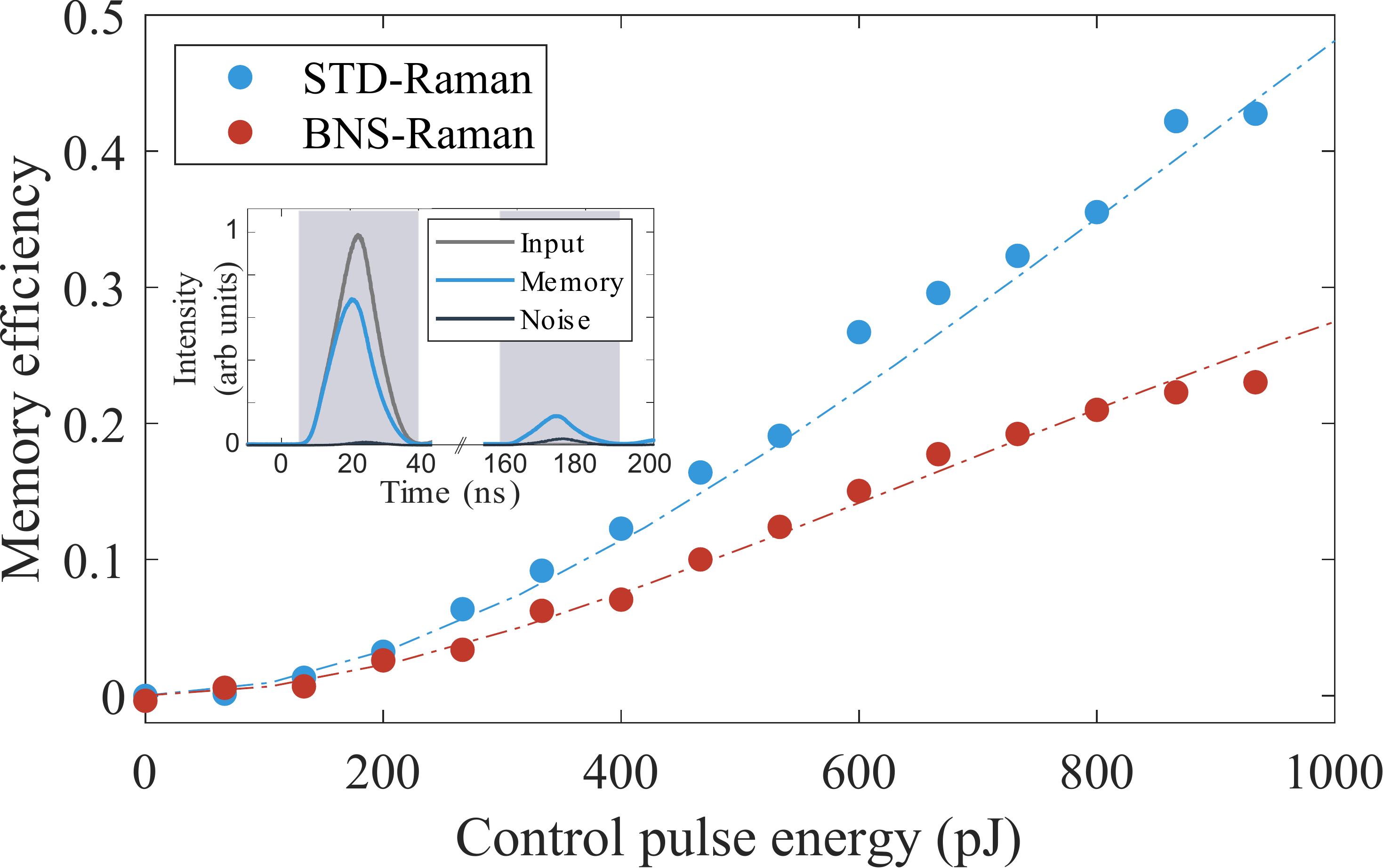}
		\caption{{Memory efficiency as a function of control pulse energy. The error bars, estimated from Poissonian errors on the number of detection events, are within the marker size. The dashed lines are the theoretical predictions from our numerical simulations. The inset shows typical arrival time histograms for the input signal (control pulses off), the memory interaction (signal and control pulses), and noise (control only, no signal). The shaded grey regions show the integration windows which are set to 35~ns. The memory retrieval time is 150~ns, and the input photon number for the STD- (\magic-) Raman memory is 3.5 (3.2).} \label{fig:MemTraces}}
	\end{figure}

	Figure~\ref{fig:MemTraces} shows the memory efficiency as a function of control pulse energy in these two regimes, and a typical arrival time histogram of the memory interaction measured on the single photon avalanche photodiodes (APDs). The memory efficiency is defined as the ratio of the energy of the retrieved signal to the input signal, and in the STD memory includes contributions from the desired memory interaction as well as FWM gain at high coupling strengths~\cite{Thomas2017}. The memory efficiency is lower in the \magic-Raman case since we have suppressed the four-wave mixing gain process. The memory lifetime is $(294\, \pm\, 26)$~ns for the STD-Raman memory, and $(625\, \pm\, 36)$~ns for the \magic-Raman memory. The factor limiting these timescales is the extinction ratio that can be achieved with the intensity modulator switching the optical pumping light, where any residual leakage during storage continues to pump the atoms and causes the spin-wave to be depleted. This extinction ratio varies over time which causes the discrepancy in the storage time between the two experiments. A better modulator would allow memory lifetimes limited in this case by the diffusion of atoms out of the control beam, thus similar to the previously demonstrated 2~$\mu$s lifetime~\cite{Michelberger2015}.
	
\begin{figure}
		\includegraphics[width=\linewidth]{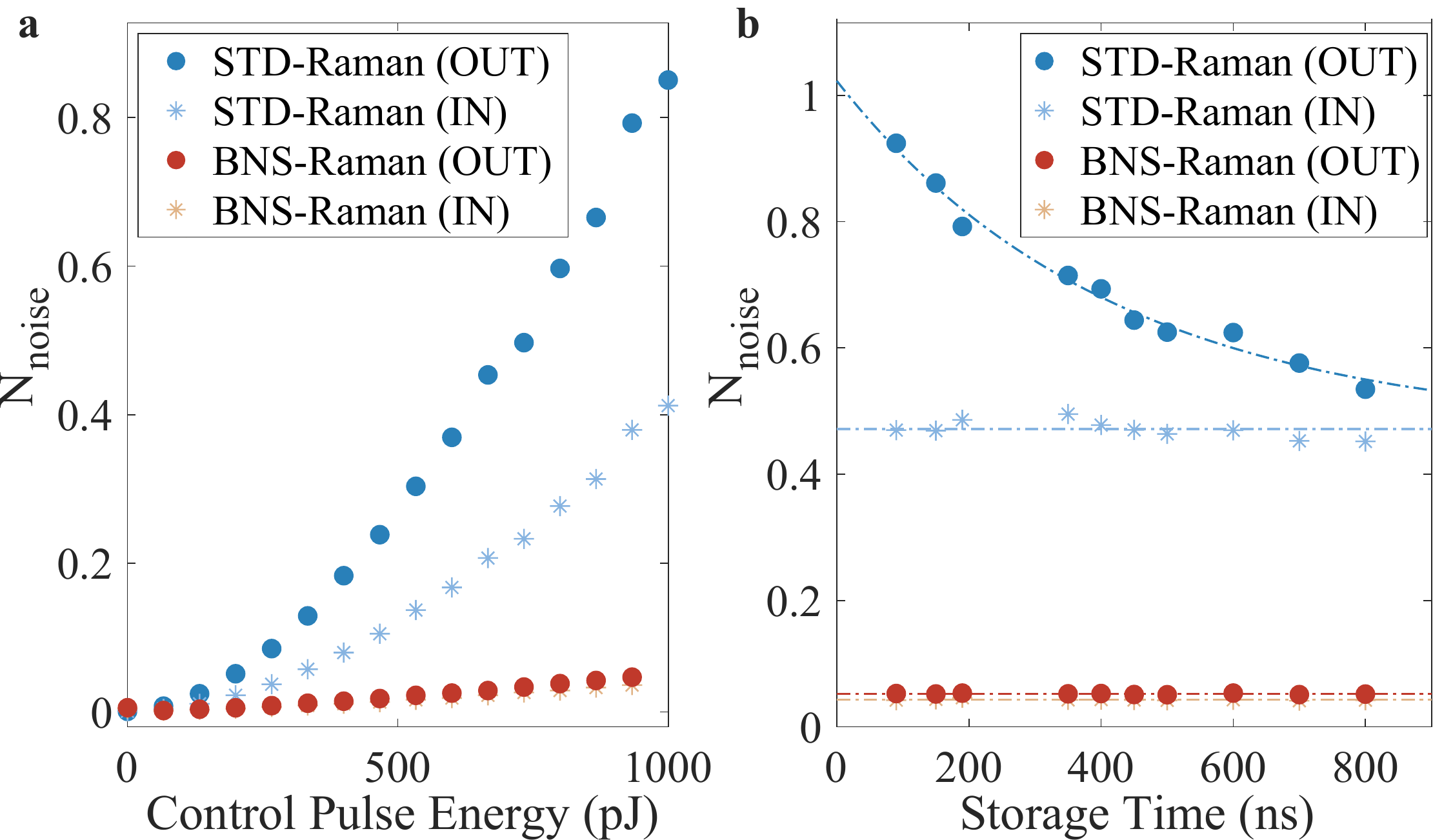}
		\caption{{The measured noise photons per pulse, $N_\mathrm{noise}$, in the input and storage time-bins as a function of (a) control pulse energy and (b) memory storage time for the \magic\, (red) and STD (blue) Raman memory. The storage time for (a) is 70~ns, and the control pulse energy for (b) is 930~pJ. The error bars, estimated from Poissonian errors on the number of detection events, are within the marker size.}  \label{fig:NoiseVs}}
	\end{figure}

	To confirm the noise suppression we measure the average number of noise photons generated per control pulse, $ N_\mathrm{noise} $, as shown in Figure~\ref{fig:NoiseVs}. The amount of noise is strongly suppressed in the \magic-Raman scheme. At a control pulse energy of 930~pJ and storage time of 70~ns the unconditional noise level is reduced from $N_\mathrm{noise}^\mathrm{(STD)} = 0.793(2)$ to $N_\mathrm{noise}^\mathrm{(\magic)} = 0.0467(5) $ photons per pulse  -- a decrease by a factor of 17. The memory efficiency for these parameters is $ \eta^\mathrm{(STD)} = 42.8(2) \%$ and $ \eta^\mathrm{(\magic)} = 23.0(1) \%$  respectively and therefore the noise-to-efficiency ratio, $\mu_1 = N_\mathrm{noise}/ \eta$, sees a significant decrease from $\mu_1^\textrm{(STD)} = 1.85(9)$ to $\mu_1^\textrm{(\magic)} = 0.20(2)$. This demonstrates that the \magic-Raman scheme is a powerful, technically simple method to reduce the noise in the Raman memory, without detriment to the memory efficiency or lifetime. 
	
	We gain further insight by looking at how the noise level scales with the memory readout time. For the STD-Raman memory the noise in the retrieval time bin decreases with storage time with a decay constant of $(380 \pm 36)$~ns. This indicates that the noise process involves an atomic coherence which decays at a similar rate as the memory efficiency --  consistent with four-wave mixing noise. In contrast, for the \magic-Raman memory the noise is independent of the read-out time. We also see that in the STD-Raman memory the noise is significantly higher in the retrieval time-bin than in the input time-bin. This is again consistent with SFWM noise which builds up with subsequent applications of the control pulse due to the generation and partial retrieval of a spin-wave excitation with each pulse~\cite{Michelberger2015}. Other sources of noise such as spontaneous scattering due to imperfect optical pumping and fluorescence would generate the same amount of noise photons on every application of the control pulses and would therefore be equal in all time-bins. For the \magic-Raman memory the noise is almost identical for the storage and retrieval time-bins. These results give a strong indication that we have suppressed the four-wave mixing process, and, for a control pulse energy of 930~pJ, have residual noise sources remaining of $\sim 0.05$ photons per pulse.
	
	\emph{Noise Characterisation} -- To further characterise the output of the memory and confirm the suppression of the SFWM noise, we consider the second order autocorrelation function of the output photonic state, $\gtwo_\mathrm{out}$, which allows us to investigate the different noise sources in more detail. This is defined as~\cite{Christ_2011}:

	\begin{eqnarray}
	\gtwo_\mathrm{out}(\tau) & = & \frac{\int \mathrm{d} t \, \left \langle S^\dagger_\mathrm{out}(t)  S^\dagger_\mathrm{out}(t+\tau)  S_\mathrm{out}(t+\tau)  S_\mathrm{out}(t)\right \rangle }{\left( \int \mathrm{d} t \, \left \langle  S^\dagger_\mathrm{out}(t)S_\mathrm{out}(t) \right \rangle \right) ^2},\nonumber
	\end{eqnarray}

	\noindent where $S_\mathrm{out}$ is the output signal mode operator, and $\gtwo(\tau=0) < 1$ signifies a field with non-classical statistics. We consider the average number of noise photons arriving at the detector for three noise processes that could contribute to the photon number statistics: (1) Spontaneous Raman Scattering, $N_\mathrm{SRS}$, either from FWM or from spontaneous Stokes scattering from the unpumped thermal population in $\ket{3}$ with $\gtwo_\mathrm{SRS} = 2$; (2) Broadband collisional-induced fluorescence, $N_\mathrm{F}$, that is not sufficiently filtered from detection which we consider to be single-mode thermal noise with $\gtwo_\mathrm{F} = 2$; and (3) Control field leakage, $N_\mathrm{L}$, which we consider to be zero evidenced by measuring detection dark counts when the memory medium is removed. We derive an expression for the $\gtwo(0)$ of the retrieved state as a function of the output photon number $N_\mathrm{out} = \eta N_\mathrm{in}$ taking into account these different noise processes, and predict:
	
	\begin{eqnarray}
	\gtwo_\mathrm{out} & = & 1 + \frac{a N_\mathrm{out}^2 + 2 N_\mathrm{SRS} N_\mathrm{out} + N_\mathrm{SRS}^2 + N_\mathrm{F}^2 }{( N_\mathrm{out} + N_\mathrm{SRS}+N_\mathrm{F})^2}  \label{eqn:g2VsNin} 
	\end{eqnarray}
	
	\noindent where $a = \gtwo_\mathrm{in} \mathcal{G}_{ss}/\eta^2 -1$, and $\mathcal{G}_{ss}$ is the integrated Green's function kernel describing the linear mapping from the input signal field to the retrieved signal field. For more details see the Supplemental Material.
	
	\begin{figure}
		\includegraphics[width=\linewidth]{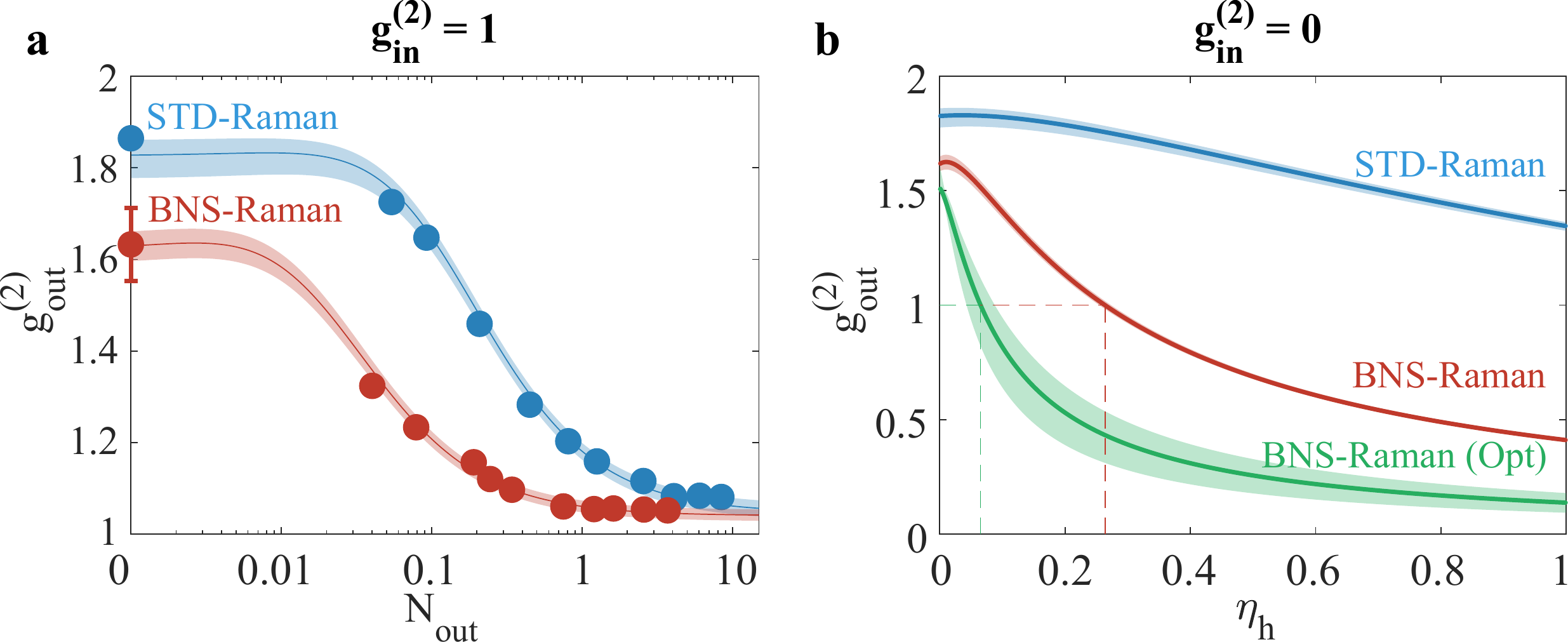}
		\caption{	{(a)~The measured $\gtwo$ of the retrieved state from the memory as a function of the retrieved photon number, $N_\mathrm{out}$ for the \magic\, and STD Raman memory. The control pulse energy is 330~pJ and the storage time is 150~ns. The solid lines are the fit to data using Equation~\ref{eqn:g2VsNin}, and the shaded regions indicate the 95\% confidence intervals on the fit. (b)~The predicted $\gtwo_\mathrm{out}$ for a single photon input with $g^{(2)}_\mathrm{in} = 0$ as a function of the input photon number or heralding efficiency, $\eta_h$, using the fitted parameters. The green line indicates the $\gtwo$ that could be reached by optimising the BNS scheme via improvement of the optical pumping. }  \label{fig:g2VsNin}}
	\end{figure}
	
	To quantify the relative contributions of these different noise sources, we measure the second-order autocorrelation of the retrieved optical field for coherent state inputs $\gtwo_\mathrm{in} = 1$ with average photon numbers varying from $N_\mathrm{in} \sim 0.5$ to 80, as well as measuring the autocorrelation of the noise with $N_\mathrm{in}=0$. The output signal is sent to a Hanbury-Brown-Twiss set-up comprised of a half waveplate, a polarising beam splitter, and two fibre-coupled APDs. The correlations between the two detectors are measured using a time-tagger (Swabian Instruments Time Tagger 20) to calculate $\gtwo_\mathrm{out}$. 
	
	Figure~\ref{fig:g2VsNin}(a) shows the results for the measured $\gtwo_\mathrm{out}$ as a function of the output photon number, $N_\mathrm{out} = \eta N_\mathrm{in}$, in both memory configurations, for a control pulse energy of 330~pJ and a storage time of 150~ns. The measured $\gtwo_\mathrm{out}$ for the \magic-Raman is lower than that of the STD-Raman for all input photon numbers tested. Further, the measured $\gtwo_\mathrm{out}$ approaches unity more rapidly for the BNS-Raman as the input photon number is increased. Fitting Equation~\ref{eqn:g2VsNin} to these data allows us to estimate the relative spontaneous Raman scattering and fluorescence noise contributions. $N_\mathrm{SRS}$ is reduced from $N_\mathrm{SRS}^\mathrm{(STD)}=81(2)\times10^{-3}$ to $N_\mathrm{SRS}^\mathrm{(BNS)}=11.0(5)\times10^{-3}$ photons per pulse, with $N_\mathrm{F}$ decreasing slightly from $N_\mathrm{F}^\mathrm{(STD)}= 9(3)\times10^{-3}$ to $N_\mathrm{F}^\mathrm{(BNS)}=3.8(5)\times10^{-3}$. Therefore, along with the significant decrease in the average noise, the \magic\, case presents a change in the SRS-to-fluorescence ratio ($\sim3$, compared to $\sim9$ for the Raman memory) resulting in a noticeable reduction of $\gtwo_\mathrm{out}$ at zero input, whilst the increased SNR allows for faster scaling to $\gtwo_\mathrm{in}$ as the input photon number is increased.
	
	To distinguish whether the remaining noise from spontaneous Raman scattering, $N_\mathrm{SRS}$, is due to four-wave mixing or Stokes scattering due to imperfect optical pumping, we measure the noise as a function of the optical pumping efficiency. We find that the noise decreases linearly with the amount of residual population in the storage state, which is consistent with noise from spontaneous scattering from the unpumped population. We measure that we can decrease the total noise to $N_\mathrm{noise} = 6.6(2) \times 10^{-3}$ with higher optical pumping power (see Supplemental Material). Even for perfect optical pumping we predict $N_\mathrm{noise} = 5(2) \times 10^{-3}$, which agrees well with the extracted value of $N_F^\mathrm{(BNS)}=3.8(5)\times10^{-3}$. We therefore conclude that the residual noise is a combination of fluorescence noise and spontaneous Raman scattering due to imperfect optical pumping, and that we have successfully eliminated SFWM noise.

	To explore the efficacy of this scheme for enabling quantum-level storage, we use Equation~\ref{eqn:g2VsNin} together with the fitting parameters from the weak coherent state data to predict the output photon statistics for the case when the input is a single photon Fock state with $\gtwo_\mathrm{in} = 0$. Figure~\ref{fig:g2VsNin}(b) shows the predicted $\gtwo_\mathrm{out}$ as a function of the probability for a single photon to arrive at the input of the quantum memory, $\eta_\mathrm{h}$. The STD-Raman memory is unable to produce a non-classical output state, even for unit probability, due to the significant noise contribution. In contrast, the BNS-Raman case with the large reduction FWM noise is able to output nonclassical states for heralding efficiencies exceeding $(26.4 \pm 0.5)\,\%$. This is well within the performance parameters of existing technologies, with heralding efficiencies as high as 87\% possible~\cite{Kaneda:16}. 
	
	By measuring the decrease in noise as we increased the optical pumping power we have determined that we could reduce the total noise to $N_\mathrm{noise} = 6.6(2) \times 10^{-3}$, of which approximately half is due to fluorescence ($N_\mathrm{F}^\mathrm{(BNS)}=3.8(5)\times10^{-3}$) and half is due to imperfect optical pumping. Furthermore, improved pump switching extinction will prevent spin-wave depletion during storage, improving our efficiency from 10.2\% to 12.7\% for this control pulse energy, with the same level of noise. These changes would give $\mu_1^\mathrm{opt} = 0.052(3)$, or an upper bound on the conditional fidelity of a retrieved qubit of around $\mathcal{F} = 0.95$~\cite{Gundogan2015}. This would also yield the green line in Figure~\ref{fig:g2VsNin}(b), or a drop in the requisite single-photon heralding efficiency for non-classical read-out to $(6.5 \pm 0.6)\,\%$. In this case a deterministic single photon source could be retrieved from the memory with $\gtwo_\mathrm{out} = 0.14 \pm 0.02$, significantly below the non-classical threshold. Further improvement of the optical pumping, and better spectral filtering to remove fluorescence noise would allow for even more dramatic improvements.

	\emph{Conclusion} -- We have demonstrated a novel scheme for the suppression of noise in a quantum memory in a warm atomic vapour by means of coherent destructive interference of SFWM and absorption. We have shown that this built-in noise suppression offers reduction in SFWM noise in a Raman memory by more than an order of magnitude, reaching a level where non-classical operation is possible. By quantifying different noise contributions in a quantum memory in terms of underlying physical processes, we conclude that the remaining noise is a combination of collisional-induced fluorescence and spontaneous Raman scattering from imperfect optical pumping. Further, we estimate that this technical noise could be suppressed by improved frequency filtering and by increasing the optical pumping efficiency. Our scheme is broadly applicable to any off-resonant memory protocol or any system that suffers from SFWM noise, and in particular this method would be efficacious in a cold atomic system where collisional induced fluorescence noise is negligible. This is a technically simple and effective noise suppression scheme, that paves the way towards quantum level storage in a long-lived, broadband, room-temperature quantum memory. \\
	
	{\emph{Acknowledgments} -- 	We thank Khabat Heshami, Oscar Lazo-Arjona and Jonas Becker for insightful discussions. This work was supported by the UK Engineering and Physical Sciences Research Council (EPSRC) through the Standard Grant No. EP/J000051/1, Programme Grant No. EP/K034480/1, the EPSRC Hub for Networked Quantum Information Technologies (NQIT), and an ERC Advanced Grant (MOQUACINO). This work has received funding from the European Union's Horizon 2020 Research and Innovation Program under grant agreement No. 665148 (QCUMbER). SET and JHDM are supported by EPSRC via the Controlled Quantum Dynamics CDT under Grants No. EP/G037043/1 and No. EP/L016524/1, and TMH is supported via the EPSRC Training and Skills Hub InQuBATE Grant EP/P510270/1. JN acknowledges financial support from a Royal Society University Research Fellowship, DJS acknowledges financial support from an EU Marie Curie Fellowship No. PIIF-GA-2013-629229 and PML acknowledges financial support from a European Union Horizon 2020 Research and Innovation Framework Programme Marie Curie individual fellowship, Grant Agreement No. 705278 (Quantum BOSS). }
	
	
	\bibliography{BNS_Raman}
	\clearpage
		\appendix
		\setcounter{figure}{0} \renewcommand{\thefigure}{A.\arabic{figure}}
		\setcounter{equation}{0} 
	
		\small 
		\section{Equations of Motion and Numerical Simulations}

		The equations of motion for the signal field, $\hat{S}$, and the anti-Stokes field, $\hat{A}$, interacting with an ensemble of atoms via a strong control pulse with Rabi frequency $\Omega$, and generating a spin-wave excitation $\hat{B}$ are given by~\cite{Nunn2017}:
		
		\begin{eqnarray}
			\left (c\partial_z+\partial_t \right ) \hat{S} & = & \im c \sqrt{\frac{{d} \gamma}{L}}\frac{\Omega}{\Gamma_s} \left( \hat{B} + \hat{F}_s \right) - \kappa_s \hat{S} \label{eqn:raman_EqOM_S} \\
			\left (c\partial_z+\partial_t \right ) \hat{A} & = & \im c \sqrt{\frac{{d} \gamma}{L}}\frac{\Omega}{\Gamma_a} \left( \hat{B}^\dagger + \hat{F}_a \right)  - \kappa_a \hat{A} \label{eqn:raman_EqOM_A} \\
			\partial_t  \hat{B} & = & -\im \Omega^*\sqrt{\frac{{d} \gamma}{L}} \left( \frac{1-\alpha}{\Gamma_s}+\frac{\alpha}{\Gamma_s^*} \right) \hat{S}  \nonumber \\
			&& + \im \Omega^* \sqrt{\frac{{d} \gamma}{L}} \left( \frac{1-\alpha}{\Gamma_a} + \frac{\alpha}{\Gamma^*_a} \right)\hat {A}^\dagger   \nonumber \\
			&& - \left( \frac{1}{\Gamma^*_a} +\frac{1}{\Gamma_s} \right  ) |\Omega|^2 \hat{B} - \Omega^* \left( \frac{\hat{F}_a^\dagger}{\Gamma_a^*} + \frac{\hat{F}_s}{\Gamma_s} \right) . \nonumber \\
			&& \label{eqn:raman_EqOM_B}
		\end{eqnarray}
		
		\noindent Here $\Gamma_{s,a} = \gamma + \im \Delta_{s,a} $ is the complex detuning of the signal and anti-Stokes fields, and $\gamma = \gamma_N + \gamma_P$ is the total Lorentzian linewidth of the excited state including the natural linewidth, $\gamma_N$, and pressure broadening due to collisions with the buffer gas, $\gamma_P$. ${d}$ is the pressure-broadened optical depth of the ensemble, which is related to the on-resonance optical depth, $d_0$, by ${d} = d_0 \gamma_{N} /\gamma$, and $\alpha$ is the proportion of atoms that remain in state $\ket{3}$ due to finite optical pumping efficiency. $\hat{F}_{s,a}$ are the Langevin noise operators which are introduced in addition to the decay terms on the atomic coherences, described by $\Gamma_{s,a}$, to account for fluctuations and ensure that the bosonic commutation relations still hold. These operators will only appear in normally ordered expectation values in expressions for the memory efficiency, and since the operators are initially in the vacuum state these expectation values are zero and we neglect such terms in our numerical simulations. However, if there is significant occupation of the excited state due to linear absorption of the control field, it may be that this assumption is no longer valid, and hence our model may not capture all sources of noise.
		
		We solve Equations~\ref{eqn:raman_EqOM_S}-\ref{eqn:raman_EqOM_B} numerically using a combination of the Runge-Kutta method and Chebyshev iteration method, and consider three cases:
		
		\begin{enumerate}
			\item \textbf{\magic \, Raman}. We consider a range of detunings around the absorption condition $\Delta_s = -2\Delta_\mathrm{hf}$.
			\item \textbf{STD Raman}. A similar range of detunings as in (1) is used for the blue-detuned case $\Delta_s = +2\Delta_\mathrm{hf}$. 
			\item \textbf{Ideal Raman} We artificially turn off the four-wave mixing process and calculate the memory efficiency for the ideal, red-detuned Raman memory.
		\end{enumerate}
		
		
		\begin{figure}
			\centering
			\includegraphics[width=0.95\linewidth]{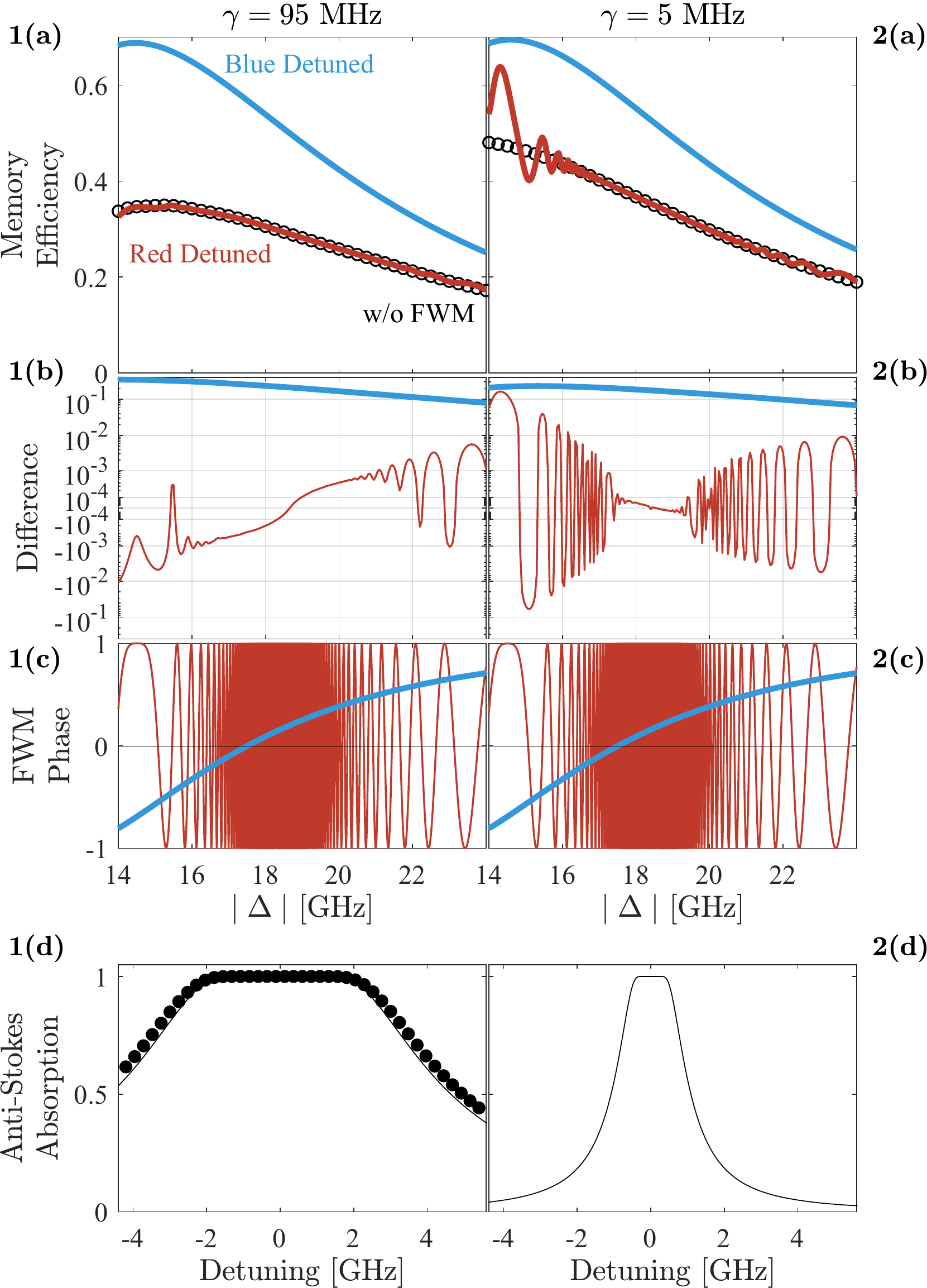}
			\caption{\footnotesize 	{The left-hand column, 1(a)-(d), is for the relevant case of Cs with a buffer gas, $\gamma = 95\,\mathrm{MHz}$; the right-hand column, 2(a)--(d), is without a buffer gas, $\gamma = \gamma_\mathrm{N}$. Note that plots (a)--(c) have the absolute detuning $\mid \Delta \mid$ (a) Numerical simulations of the memory efficiency for three cases, STD-Raman (`Blue Detuned' -- blue line), \magic-Raman (`Red Detuned' -- red line), and ideal Raman without four-wave-mixing in the equations (`w/o FWM' -- black circles). (b) Difference between blue/red detuned memory efficiency and `w/o FWM' memory efficiency in (a). (c) Four-wave mixing phase for the blue and red detuned cases. (d) Anti-Stokes absorption. Note that the centre of the feature is at zero detuning. For 1(d) we include the absorption measured with a narrowband laser (black dots).} \label{fig:Simulation}}
		\end{figure}

		The results of these simulations for an on-resonance optical depth of $ d_0 = 2.9 \times 10^4$ and a control pulse energy of 750~pJ are shown in Figure~\ref{fig:Simulation}, both for a pressure broadened linewidth of $\gamma = 96$~MHz (left) and the natural linewidth $\gamma = \gamma_N = 5.2$~MHz (right). We see that the efficiency in case (2) is higher than the ideal case (3) due to four wave mixing gain. The efficiency for the BNS-Raman (case 1) is almost identical to the ideal case (3) over a broad range of detunings around the condition $\Delta = -2\Delta_\mathrm{hf} = -18.4$~GHz due to the built-in noise suppression. 
		
		The second panel in Figure~\ref{fig:Simulation} shows the difference in efficiency between cases (1) and (3) (red), and cases (2) and (3) (blue), on a log scale, and quantifies the increase in efficiency due to four-wave mixing gain. We see that the four-wave mixing gain is significantly suppressed in the \magic  \, scheme over the whole 10~GHz frequency range, and at the exact absorption condition we see a suppression in the gain process by over four orders of magnitude compared to the standard Raman memory. 
		
		We note that the efficiency in case (1) oscillates around the ideal case (seen more clearly for the case with no pressure broadening) and this is due to the four-wave mixing phase-matching condition. The phase mismatch, $\delta k = 2 k_c - k_s - k_a$, is shown in the third panel in Figure~\ref{fig:Simulation}, and we see that it is rapidly changing due to the strong absorption feature at the anti-Stokes frequency. The four-wave mixing process goes in and out of phase, resulting in an energy transfer back and forth between the four fields. We note that the dispersive feature is very broad, and the four-wave mixing process is poorly phase-matched over a wide range of detunings around $\Delta = -2 \Delta_\mathrm{hf}$. This allows suppression of SFWM noise using atomic absorption even for a broadband noise field, and therefore this BNS-suppression scheme is widely applicable for narrow or broadband memory protocols.

		\section{Experimental Set-up}
		
		To experimentally implement the Raman memory in warm caesium vapour, we use the 6~S$_{1/2}$(F~=~3) and (F~=~4) hyperfine states as ground and storage states, $\ket{1}$ and $\ket{3}$, respectively, which are separated by $\Delta_\mathrm{hf} = 9.2$~GHz. The memory interaction is mediated by strong control pulses which drive an off-resonant two-photon Raman transition from ground to storage state via the 6~P$_{3/2}$ manifold, $\ket{2}$. The signal and control pulses are generated using a fibre-integrated electro-optic modulator to carve pulses from a continuous wave laser. The modulator features two electrodes which are driven by arbitrary waveform generators with a sampling rate of 50~Gs/s and which facilitate arbitrary phase and amplitude control of the output pulses. We generate three pulses of an intensity full-width half maximum of 10~ns to act as the signal, read-in and read-out control pulses. The signal pulse passes through a further bulk electro-optic modulator which is modulated at a frequency of 9.2~GHz to generate sidebands. The carrier frequency and blue sideband are filtered away using a Fabry-P\'{e}rot etalon, leaving the red sideband which is in two-photon resonance with the control field. The control pulses are amplified using a tapered amplifier to ensure sufficient pulse energy to drive the memory interaction. The orthogonally-polarised signal and control pulses are temporally- and spatially-overlapped, and focused to a waist radius of 130$\mu$m in the centre of a caesium vapour cell. The vapour is heated to a temperature of 83.0$^\circ$C to give a resonant optical depth of $d_0=2.98 \times 10^4$, and placed inside a $\mu$-metal magnetic shield to reduce magnetic dephasing of the spin-wave. The system is initialised by optically pumping the ensemble into the F~=~3 ground state via a counter-propagating continuous-wave laser on resonance with the $6\,\mathrm{S}_{1/2} \rightarrow 6\,\mathrm{P}_{1/2}$ transition. The EOM is switched off during the memory interaction using an EOM (EOSpace) to prevent depletion of the stored spin-wave. A buffer gas of 5 Torr of N$_2$ is mixed with the caesium vapour to allow a high pumping efficiency of $(1-\alpha) = 99.85\%$ to be reached at such high optical depths~\cite{Thomas2017}. 
		
		After the memory interaction, the strong control field is filtered away from the retrieved signal using a series of polarisation and frequency filtering. First a calcite beam displacer suppresses the control field by over 5 orders of magnitude, before the signal is coupled into a single-mode fibre. A series of Fabry-P\'{e}rot etalons (FPEs) are then used to further suppress the control field: 4 FPEs with a free-spectral range (FSR) of 18.4~GHz to maximally suppress the control frequency (9.2~GHz separated from the signal), followed by 2 FPEs with an FSR of 103~GHz, to further suppress the control field in addition to suppressing the anti-Stokes field (which is 18.4~GHz separated from the signal and therefore transmits through the first FPEs) and the broadband fluorescence noise. In total a suppression of the control field of over 110dB is achieved, while the transmission of the signal field is 15\%. After filtering, the signal is coupled into a single-mode fibre and detected with a standard fibre-coupled single-photon avalanche photo diode and a time-to-digital converter.

		\section{Derivation of $\gtwo_\mathrm{out}$}
		
		
		The equations of motion for the Raman memory are given by Equations~\ref{eqn:raman_EqOM_S}~-~\ref{eqn:raman_EqOM_B}. These equations are linear so the resulting evolution may be characterised by Green's function mappings $G_{ij}$ from initial mode $i$ to final mode $j$, which may be represented in the form:
		
		\begin{gather}
			\begin{pmatrix} \hat{S}_\mathrm{out}\\\hat{A}^\dagger_\mathrm{out}\\ \hat{B}_\mathrm{out} \end{pmatrix}
			=
			\begin{pmatrix}
				G_{ss} & G_{sa^\dagger} & G_{sb}  \\
				G_{a^\dagger s} & G_{a^\dagger a^\dagger} & G_{a^\dagger b} \\
				G_{bs}&  G_{ba^\dagger} & G_{bb}
			\end{pmatrix}
			\begin{pmatrix}
				\hat{S}_\mathrm{in} \\ \hat{A}^\dagger_\mathrm{in} \\  \hat{B}_\mathrm{in}
			\end{pmatrix}.
		\end{gather}

		We consider the input signal to be a superposition of Fock states:
		
		\begin{eqnarray}
			\ket{s_\mathrm{in}} & = & \sum_n c_n\ket{n}  \nonumber \\
			N_\mathrm{in}  & \equiv &  \bra{s_\mathrm{in}} S^\dagger S \ket{s_\mathrm{in}} \nonumber 
		\end{eqnarray}

		We take the input anti-Stokes field to be vacuum, $\ket{a_\mathrm{in}} = \ket{0_a}$, and the initial spin-wave to have population $\langle b_\mathrm{in} \rangle = \alpha$, given by number of unpumped atoms in initial state $\ket{3}$. The full input state is described as $\ket{\Psi_\mathrm{in}} = \ket{s_\mathrm{in}} \ket{0_a} \ket{b_\mathrm{in}} $ and the number of photons retrieved from the memory is given by:

		\begin{eqnarray}
			N_\mathrm{out}& =& \bra{\Psi_\mathrm{in}} S^\dagger _\mathrm{out} S _\mathrm{out}  \ket{\Psi_\mathrm{in}} \nonumber \\
			& \equiv & N_\mathrm{mem} + N^{AS}_\mathrm{SRS} + N^P_\mathrm{SRS}, \label{eqn:SS}
		\end{eqnarray}
		
		\noindent where the first term describes output photons due to the desired memory interaction, the second term describes output photons arising from spontaneous four-wave mixing, and the third term is spontaneous Raman scattering due to the initial occupation of the spin-wave from imperfect optical pumping.

		The second order autocorrelation function of the input signal is:
		
		\begin{eqnarray}
			\gtwo_\mathrm{in} &=& \frac{\bra{\psi_\mathrm{in}} S^\dagger S^\dagger S S \ket{\psi_\mathrm{in}}}{\bra{ \psi_\mathrm{in}} S^\dagger S \ket{\psi_\mathrm{in}}^2} =  \frac{\bra{\psi_\mathrm{in}} S^\dagger S^\dagger S S \ket{\psi_\mathrm{in}}}{\langle N_\mathrm{in} \rangle ^2}, \nonumber
		\end{eqnarray}
		
		\noindent and so 
		
		\begin{eqnarray}
			\bra{\psi_\mathrm{in}} S^\dagger S^\dagger S S \ket{\psi_\mathrm{in}} = \langle N_\mathrm{in} \rangle ^2 \gtwo_\mathrm{in}.\nonumber
		\end{eqnarray}

		By evaluating expressions for all the non-zero terms in $\langle S^\dagger S^\dagger S S \rangle$ we find: 
		
		\begin{eqnarray}
			\langle S^\dagger  S^\dagger S S \rangle & =&  2 \langle  S^\dagger S  \rangle^2 - N_\mathrm{in}^2(2 \eta^2 - \gtwo_\mathrm{in}\mathcal{G}_{ss}) \nonumber \\
			&& -2((N_\mathrm{SRS}^{AS})^2 - \mathcal{G}_{sa^\dagger}) - 2 ((N_\mathrm{SRS}^P)^2 - \alpha^2 \mathcal{G}_{sb}), \nonumber 
		\end{eqnarray}
		
		\noindent where $\mathcal{G}_{ij}  \equiv \iint \mathrm{d} t \, \mathrm{d}t' \, | G_{ij}(t,t')|^4$. Therefore: 
		
		\begin{eqnarray}
			\gtwo_\mathrm{out} &=&  2 -\{ N_\mathrm{in}^2 (2\eta^2 - \gtwo_\mathrm{in} \mathcal{G}_{ss}) + 2((N_\mathrm{SRS}^{AS})^2-\mathcal{G}_{sa^\dagger}) \nonumber \\
			&&  + 2 ((N_\mathrm{SRS}^P )^2 - \alpha^2 \mathcal{G}_{sb} )\}/ (\eta N_\mathrm{in}+N_\mathrm{SRS}^{AS}+N_\mathrm{SRS}^P)^2. \nonumber \\
			&& \label{eqn:g2Sig}
		\end{eqnarray}

		We  assume that the noise due to spontaneous Raman scattering would give thermal output statistics i.e. $\gtwo_\mathrm{out} (N_\mathrm{in}=0) = 2$. We hence approximate $\mathcal{G}_{sa^\dagger} = (N_\mathrm{SRS}^{AS})^2$ and $\mathcal{G}_{sb} = (N_\mathrm{SRS}^{P})^2/\alpha^2 $, which simplifies Equation~\ref{eqn:g2Sig} to

		\begin{eqnarray}
			\gtwo_\mathrm{out,sig} &=& 2 -\frac{N_\mathrm{in}^2 (2\eta^2 - \gtwo_\mathrm{in} \mathcal{G}_{ss}) )}{(\eta N_\mathrm{in}+N_\mathrm{SRS})^2}.\label{eqn:g2Sig_simp}
		\end{eqnarray}
		
		\noindent where $N_\mathrm{SRS} = N_\mathrm{SRS}^{AS} + N_\mathrm{SRS}^P$.
		
		
		We treat the noise contribution from fluorescence as an incoherent sum of fields with $N_{F}$ photons and $\gtwo_{F}$. The incoherent sum of fields 1 and 2 is given by~\cite{Michelberger2015}:
		
		\begin{eqnarray}
			\gtwo_{12}=\frac{N_1^2 \gtwo_{1} + 2N_1N_2 + N_2^2 \gtwo_{2}}{(N_1+N_2)^2}, \nonumber \label{eqn:g2add}
		\end{eqnarray}
		
		\noindent and hence
		\begin{eqnarray}
			\gtwo_\mathrm{tot}&=&1+\frac{N_S^2 - N_\mathrm{in}^2(2\eta^2 - \gtwo_\mathrm{in}\mathcal{G}_{ss})  }{N_\mathrm{tot}^2} + \frac{ N_F^2(\gtwo_{F}-1)}{N_\mathrm{tot}^2} \nonumber 
		\end{eqnarray}
		
		\noindent where photons produced in the signal mode are $N_\mathrm{S} = \eta N_\mathrm{in} + N_\mathrm{SRS}$, the total number of photons produced is $N_\mathrm{tot} = \eta N_\mathrm{in} + N_\mathrm{SRS}+ N_F$. This leads to the expression for $\gtwo_\mathrm{out}$ given in the main text:

		\begin{eqnarray}
			\gtwo_\mathrm{out} & = & 1 + \frac{a N_\mathrm{out}^2 + 2 N_\mathrm{SRS} N_\mathrm{out} + b }{( N_\mathrm{out} + N_\mathrm{SRS}+N_\mathrm{F})^2}  
		\end{eqnarray}
		
		\noindent where
		
		\begin{eqnarray}
			a &=& \gtwo_\mathrm{in} \mathcal{G}_{ss}/\eta^2 -1 \nonumber \\
			b &=& N_\mathrm{SRS}^2 + N_\mathrm{F}^2(\gtwo_\mathrm{F} -1). 
		\end{eqnarray}
				
				\begin{figure}[h]
					\centering
					\includegraphics[width=0.9\linewidth]{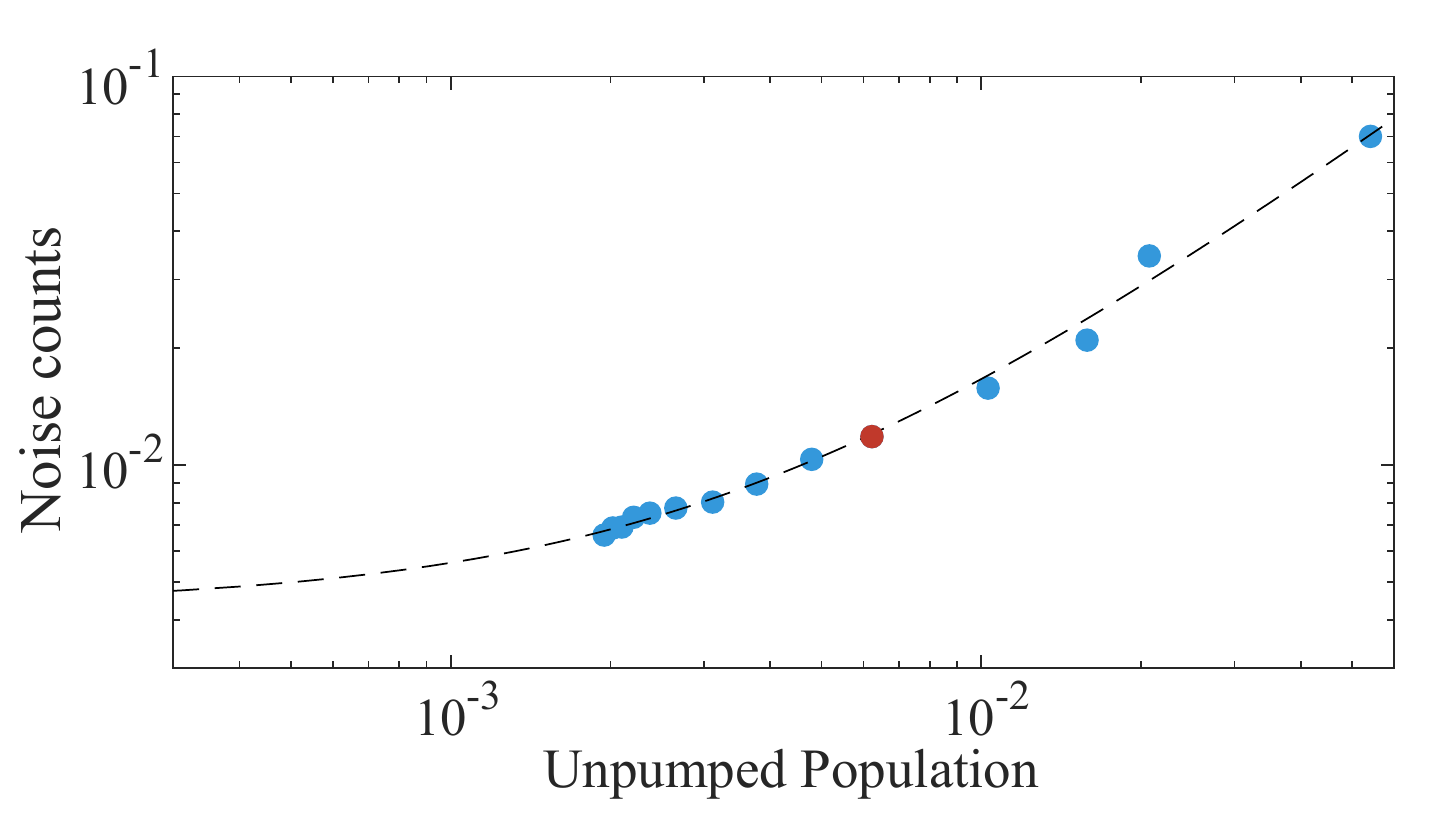}
					\caption{\footnotesize 	{Measured noise counts per pulse as a function of the proportion of the population in the storage state, $\alpha$. The red dot indicates the parameters that were used for the data presented in the main text. The black dashed line is a linear fit to the data.} \label{fig:NoiseVsPumping}}
				\end{figure}

		\section{Noise vs Optical Pumping}
		To distinguish whether the remaining noise from spontaneous Raman scattering, $N_\mathrm{SRS}$, is due to four-wave mixing or Stokes scattering due to imperfect optical pumping, we measure the average number of noise photons per pulse as a function of the pumping efficiency, $1- \alpha$. The results, shown in Figure~\ref{fig:NoiseVsPumping} indicate that the noise decreases linearly with the amount of residual population in the storage state. The control pulse energy here is 330~pJ and the storage time is 50~ns. However, these data were taken with the optical pumping beam turned on all the time, and not switching off for memory storage. Due to photo-refractive damage of the switching EOM we can  only achieve a low pumping power and a poor extinction ratio in this current demonstration. The red dot in Figure~\ref{fig:NoiseVsPumping} indicates the conditions corresponding to the measurements for the autocorrelation data in the main text. By increasing the optical pumping power to 5~mW we will be able to reduce the noise from $N_\mathrm{noise} = 11.8(2) \times 10^{-3}$ to $N_\mathrm{noise} = 6.6(2) \times 10^{-3}$. If we extrapolate these data to perfect optical pumping efficiency, $\alpha \rightarrow 0$, we find a linear offset of $N_\mathrm{noise} = ( 4.4 \pm 1.7 ) \times 10^{-3}$, which we attribute to fluorescence noise.

\end{document}